\documentclass[preprint]{aastex} 
\usepackage[squaren, Gray, cdot]{SIunits}
\usepackage{color}

\newcommand{\tab}[1]{Table~\ref{#1}}
\newcommand{\fig}[1]{Figure~\ref{#1}}

\newcommand{\sect}[1]{Section~\ref{#1}}

\newcommand{\alphamag}{\alpha_{\rm mag}}
\newcommand{\betaobs}{\beta_{\rm obs}}

\newcommand{\voltage}{\Phi}
\newcommand{\lj}{L_j}
\newcommand{\fouru}{4U~0142+61}
\newcommand{\onerxs}{1RXS~J1708--4009}
\newcommand{\onee}{1E~1841--045}

\newbox\grsign \setbox\grsign=\hbox{$>$} \newdimen\grdimen \grdimen=\ht\grsign
\newbox\simlessbox \newbox\simgreatbox \newbox\simpropbox
\setbox\simgreatbox=\hbox{\raise.5ex\hbox{$>$}\llap
     {\lower.5ex\hbox{$\sim$}}}\ht1=\grdimen\dp1=0pt
\setbox\simlessbox=\hbox{\raise.5ex\hbox{$<$}\llap
     {\lower.5ex\hbox{$\sim$}}}\ht2=\grdimen\dp2=0pt
\setbox\simpropbox=\hbox{\raise.5ex\hbox{$\propto$}\llap
     {\lower.5ex\hbox{$\sim$}}}\ht2=\grdimen\dp2=0pt
\def\simgt{\mathrel{\copy\simgreatbox}}

\newcounter{refcompteur}
  \def\generateur#1{%
  \begingroup
  \edef\next{\def\expandafter\noexpand\csname #1\endcsname{\therefcompteur}}%
  \expandafter\endgroup\next
  \addtocounter{refcompteur}{1}
}

\shorttitle{Phase-resolved X-ray spectra of magnetars} 
\shortauthors{Hascoet et al.}

\begin{document}

\title{Phase-resolved X-ray spectra of magnetars \\
and the coronal outflow model}

\author{Romain Hasco\"et\altaffilmark{1}, Andrei M. Beloborodov}
\affil{Physics Department and Columbia Astrophysics Laboratory, Columbia University, 538 West 120th Street New York, NY 10027}

\author{Peter R. den Hartog}
\affil{Stanford University HEPL/KIPAC, 452 Lomita Mall, Stanford, CA 94305-4085}

\altaffiltext{1}{\texttt{hascoet@astro.columbia.edu}}

\begin{abstract}

We test a model recently proposed for the persistent hard X-ray emission from magnetars. In the
model, hard X-rays are produced by a decelerating electron-positron flow in the closed
magnetosphere. The flow decelerates as it radiates its energy away via resonant scattering of
soft X-rays, then it reaches the top of the magnetic loop and annihilates there. 
We test the model against observations of three magnetars: \fouru, \onerxs , and \onee.
We find that the model successfully fits the observed phase-resolved spectra. 
We derive constraints on the angle between the rotational and magnetic axes of the neutron star, 
the object inclination to the line of sight, 
and the size of the active twisted region filled with the plasma flow. 
Using the fit of the hard X-ray component of the magnetar spectrum, we revisit the remaining soft X-ray component.
We find that it can be explained by a 
modified
two-temperature blackbody model.
The hotter blackbody is
consistent with a hot spot covering 
1-10\% of the neutron star surface.
Such a hot spot is expected 
at the base of the magnetospheric $e^{\pm}$ outflow,
as some particles created in the $e^{\pm}$ discharge flow back and bombard the stellar surface.

\end{abstract}


\section{Introduction}
\label{sect_intro}

Magnetars are isolated neutron stars whose emission is thought to be powered by the decay of 
intense magnetic fields \citep{duncan_1992, thompson_1996}.
Their persistent X-ray spectrum shows two peaks, 
near 1~keV and above 100~keV (e.g. \citealt{kuiper_2006, enoto_2010}).
The soft X-ray component 
likely comes from the neutron star surface. Its spectrum is modified from a simple Planck
shape by the radiative transfer in the atmosphere and magnetosphere of the star,
and it has a soft tail extending to $\sim 10$~keV where the hard component takes over.
The hard X-rays must be produced in the magnetosphere of the neutron star.

For a few magnetars, phase-resolved spectra were measured in the hard X-ray 
band  (den Hartog et al. 2008a,b). They showed 
bizarre variations with rotational phase.
Recently, \citet{beloborodov_2013} proposed a model of decelerating 
$e^{\pm}$ outflow that makes specific predictions for the hard X-ray spectrum and its
variation with rotational phase. 
In the present paper, we test the model against observations of three magnetars, 
\onee , \fouru \ and \onerxs \ (all discovered as anomalous X-ray pulsars).
The model is briefly described in \sect{model_desciption}, and 
the results of data analysis are presented in \sect{data_fit}.
We find that the model successfully fits the data and discuss implications of our results in \sect{discussion}.


\section{Coronal outflow model}
\label{model_desciption}

In the model, the hard X-ray emission is powered by the $e^\pm$ discharge near the star. 
The discharge voltage exceeds $10^9$~V and the initial Lorentz factor of created $e^\pm$ 
exceeds $10^3$ (Beloborodov \& Thompson 2007).
As the $e^{\pm}$ pairs flow out along the magnetic field lines and fill the extended 
magnetic loop, they decelerate via resonant scattering of thermal X-rays.
\citet{beloborodov_2013} showed that the outflow has two zones, adiabatic and radiative. 
The adiabatic zone is near the star where magnetic field $B>10^{13}$~G;
in this zone the resonantly scattered photons quickly convert to pairs, increasing 
the $e^\pm$ multiplicity of the flow to ${\cal M}\sim 10^2$.
The radiative zone is at higher altitudes where $B <10^{13}$~G; here the resonantly 
scattered photons escape and the outflow enters the radiative regime,
losing its kinetic energy.

The interaction of radiation and $e^\pm$ outflow was 
studied in detail
in \citet{beloborodov_2013b}.
It was shown that the outflow Lorentz factor $\gamma_\pm$ decreases proportionally 
to the local magnetic field,
\begin{equation}
   \label{eq_gpm}
   \gamma_{\pm} \approx 10^2 \frac{B}{B_Q} \, .
\end{equation}
The  average energy of resonantly scattered photons is given by
\begin{equation}
\label{eq_esc}
  E_{\rm sc} \approx \gamma_{\pm} \frac{B}{B_Q} m_e c^2 \approx 5 
    \, \gamma_\pm^2
  \  \mathrm{keV}.
\end{equation}
Equations~(\ref{eq_gpm}) and (\ref{eq_esc}) imply that the
spectrum of escaping hard X-rays has an average photon index $\Gamma=1.5$
and its luminosity peaks at the high-energy end \citep{beloborodov_2013}.
The spectrum cuts off in the MeV band, because the multi-MeV photons are 
emitted in the adiabatic zone and cannot escape.
The predicted spectral shape, however, strongly depends on the viewing angle,
and may significantly differ from the average spectrum.

For a given position of the observer and location of the active magnetic loop 
the model predicts a well-defined phase-dependent radiation spectrum
that can be quantitatively compared with observations.
The spectrum can be calculated using the Monte-Carlo technique as described in \citep{beloborodov_2013}. 
We have developed a different, independent code that performs a similar calculation  
using direct integration of the angle-dependent emissivity over the j-bundle.
This method does not include photon splitting, which is important at high energies.
We have tested the code against the Monte-Carlo results presented in \citet{beloborodov_2013} 
and found excellent agreement at photon energies below 400~keV.

In this paper, we assume that the magnetosphere is axisymmetric and close to the dipole 
configuration in the region of interest (where $B<10^{13}$ G). The dipole moment of the 
neutron star $\mu_{\rm m}$ is provided by  measurements of its spindown rate; it is taken 
from 
\cite{dib_2007} for \fouru, and from \citet{dib_2008} for \onerxs \ and \onee.
We also assume that the active, current-carrying region of the magnetosphere (``j-bundle''),
which is filled with the $e^\pm$ flow, is axisymmetric.
Then the model is completely defined by four remaining parameters:
\begin{itemize}
\item $\theta_{j}$, the half opening angle (from the magnetic axis) of the j-bundle footprint 
on the star.
\item $\lj$, the bolometric luminosity radiated by the j-bundle.
\item $\alphamag$, the angle between the rotation axis and magnetic axis of the neutron star.
\item $\betaobs$, the angle between the rotation axis and the observer line of sight.
\end{itemize}
In addition, the point of zero rotational phase must be kept as a free parameter
to fit phase-resolved data.
The parameter $\lj$ is related to two theoretical quantities --- the discharge voltage $\voltage$ 
and the twist angle of the magnetic field lines $\psi$ (Beloborodov 2009), 
\begin{equation}
\label{eq_luminosity}
    \lj \simeq 2 \times 10^{35} \   \psi \left( \frac{\voltage}{10^{10} \ \mathrm{V}} \right)  \left( \frac{\mu_{\rm m}}{10^{32} \ \mathrm{G \ cm^3}} \right)  
\left( \frac{R}{10 \ \mathrm{km}} \right)  \left(\frac{\theta_{j}}{0.3}\right)^4 \ 
\mathrm{erg~s}^{-1} \, ,
\end{equation}
where 
$R$ is the radius of the neutron star, which we fix at 10~km. A twist angle $\psi\sim 1$
is expected for a strongly twisted magnetosphere;
measured values of $\lj$ 
and $\theta_j$ can provide an order-of-magnitude for the discharge voltage $\voltage$.

\begin{figure}[h]
\begin{center}
\begin{tabular}{c}
\includegraphics[width=0.5\textwidth]{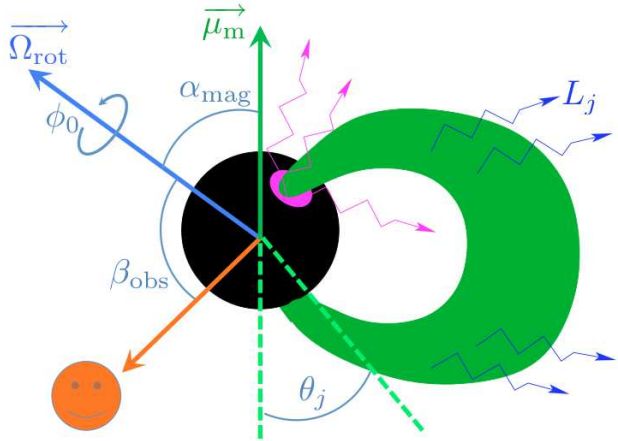} 
\end{tabular}
\end{center}
\caption{
Schematic picture of the coronal outflow model indicating its parameters (see text).
The current-carrying
magnetic loop (``j-bundle'', shown in green) is filled by 
the relativistic $e^{\pm}$ flow launched from the discharge zone near the neutron star.
At high altitudes where $B<10^{13}$~G, the $e^\pm$ outflow converts its kinetic energy to
hard X-rays (blue arrows) via resonant scattering of thermal X-rays.
Some particles from the $e^\pm$ discharge flow toward the star and bombard it,
forming a hot spot at the footprint of the j-bundle (magenta).
}
\label{fig_cartoon}
\end{figure}

The j-bundle footprint can form a hot spot on the neutron star, as some 
of the relativistic particles created in the $e^\pm$ discharge flow back to the stellar 
surface and bombard it. 
This motivates us to consider a two-temperature model for the soft X-ray emission from 
the stellar surface, with the hotter blackbody associated with the j-bundle footprint.
The expected area of the hot spot is related to $\theta_j$ by
\begin{equation}
\label{eq_hotspot_area}
   \mathcal{A}_{\rm h} \sim \frac{1}{4} \theta_j^2 \mathcal{A}_{\rm ns} \approx 0.02 \left(\frac{\theta_j}{0.3}\right)^2 \mathcal{A}_{\rm ns} \, ,
\end{equation}
where $\mathcal{A}_{\rm ns}=4\pi R^2$ is the surface area of the neutron star.
The model is illustrated in \fig{fig_cartoon}.


\section{Spectral fits}
\label{data_fit}

For \fouru \ and \onerxs , we used the multi-year data collected by INTEGRAL-ISGRI ($20-300$ keV), 
INTEGRAL-SPI ($10-1000$ keV), XMM-Newton  ($0.55-11.5$ keV), ASCA-GIS ($0.7-12$ keV, only for \fouru), 
RXTE-PCA ($2-60$ keV) and CGRO-Comptel (upper-limits in the range $0.75-30$ MeV, see 
\citealt{kuiper_2006}, \citealt{denhartog_2006}).
The data set for the two magnetars is the same as in \citet{denhartog_2008b, denhartog_2008}; it is described in detail there.
The data and fit by the outflow model for \onee \ are presented in \citet{an_2013}; 
below we show the results of An et al. (2013) 
for comparison with \fouru \ and \onerxs.

\subsection{Hard component}
\label{fit_hard_component}

We first focus on the data above 10~keV, where the relativistic outflow 
dominates the observed emission. 
In this energy range, we search for a combined fit of the phase-averaged spectrum 
of the total (pulsed + unpulsed) emission and three phase-resolved spectra of the 
pulsed emission. The three phase bins are the same as in
\citet{denhartog_2008b, denhartog_2008}.

The entire parameter space of the model (\sect{model_desciption}) was discretized on a grid, 
and we calculated $\chi^2$ and the corresponding $p$-value for the model at each grid point. 
The obtained map of $\chi^2$ 
provided a reliable way to find the best fit, avoiding the risk of converging to a local 
minimum with an optimization algorithm.
We found that $\chi^2$ is a rather complex function of 
the model parameters and standard optimization software such as XSPEC would fail to identify the best fit. 
Remarkably, we found that $\chi^2$ has a sharp minimum in a well
localized region of the parameter space.
The best fit parameters for each magnetar are summarized in \tab{tab_hard_comp}.

\begin{figure}[t]
\begin{center}
\hspace*{-1.cm}
\begin{tabular}{cccc}
$p$-value scale & \onee & \fouru & \onerxs \\
\includegraphics[trim = 9cm 0.1cm 4.5cm 0cm, width=0.1\textwidth]{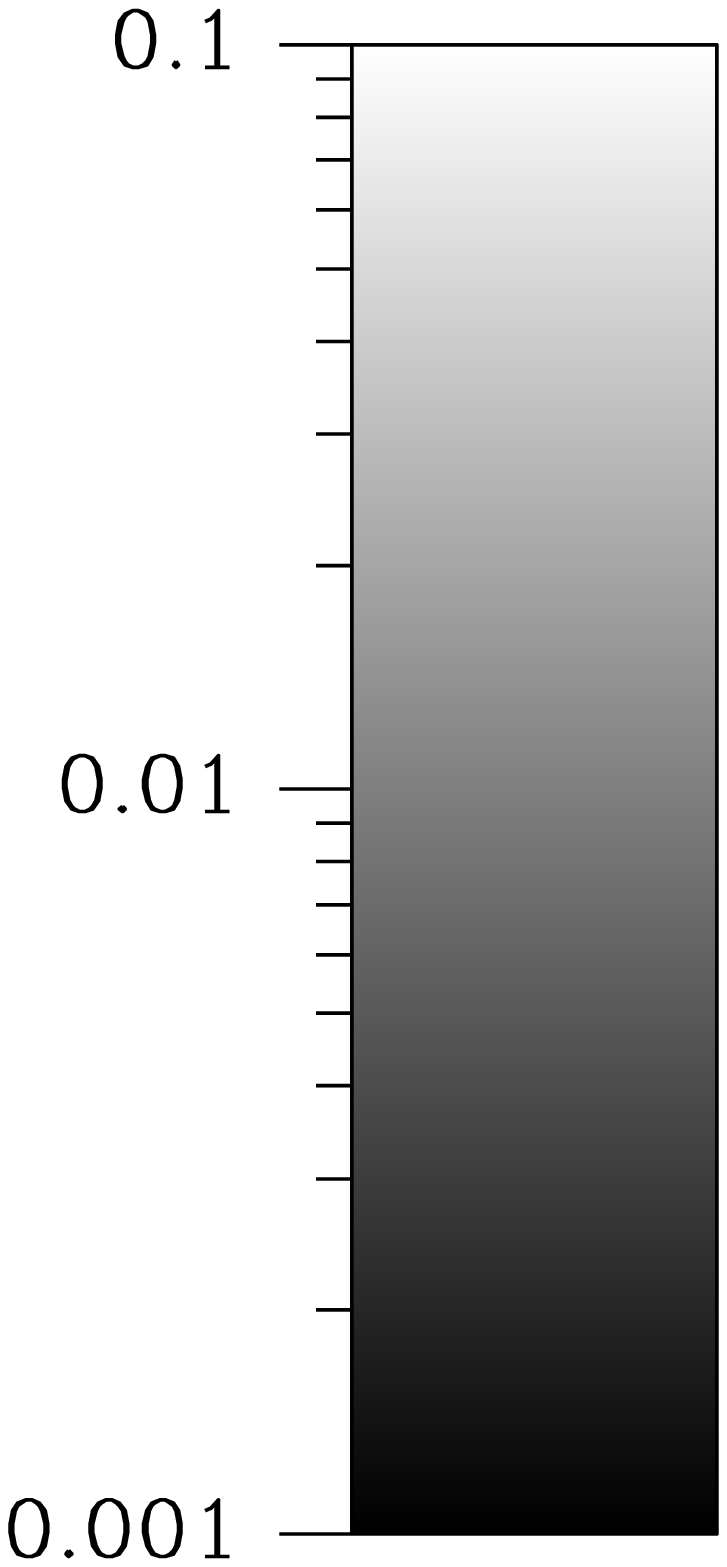} & 
\includegraphics[width=0.3\textwidth]{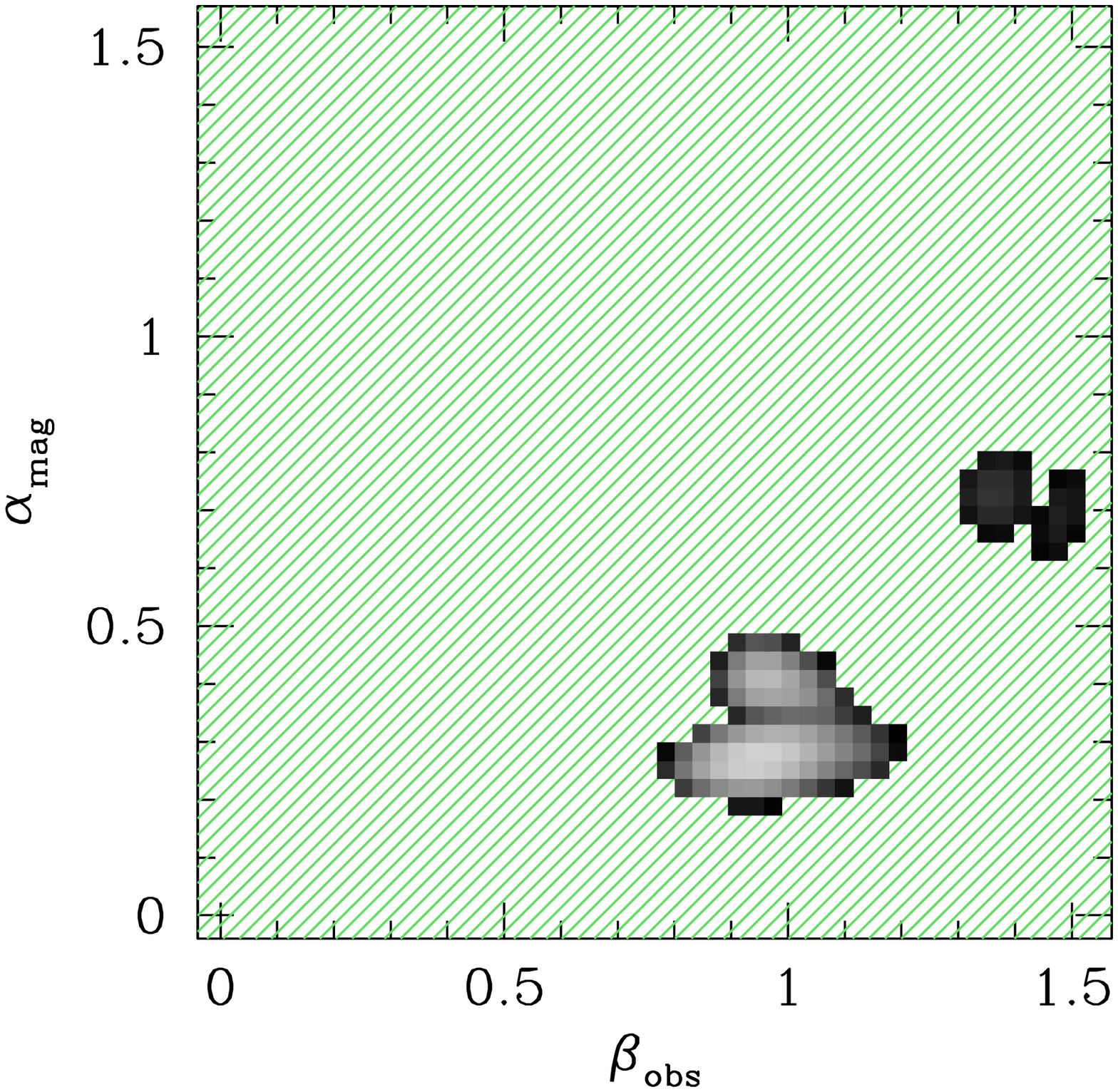} & 
\includegraphics[width=0.3\textwidth]{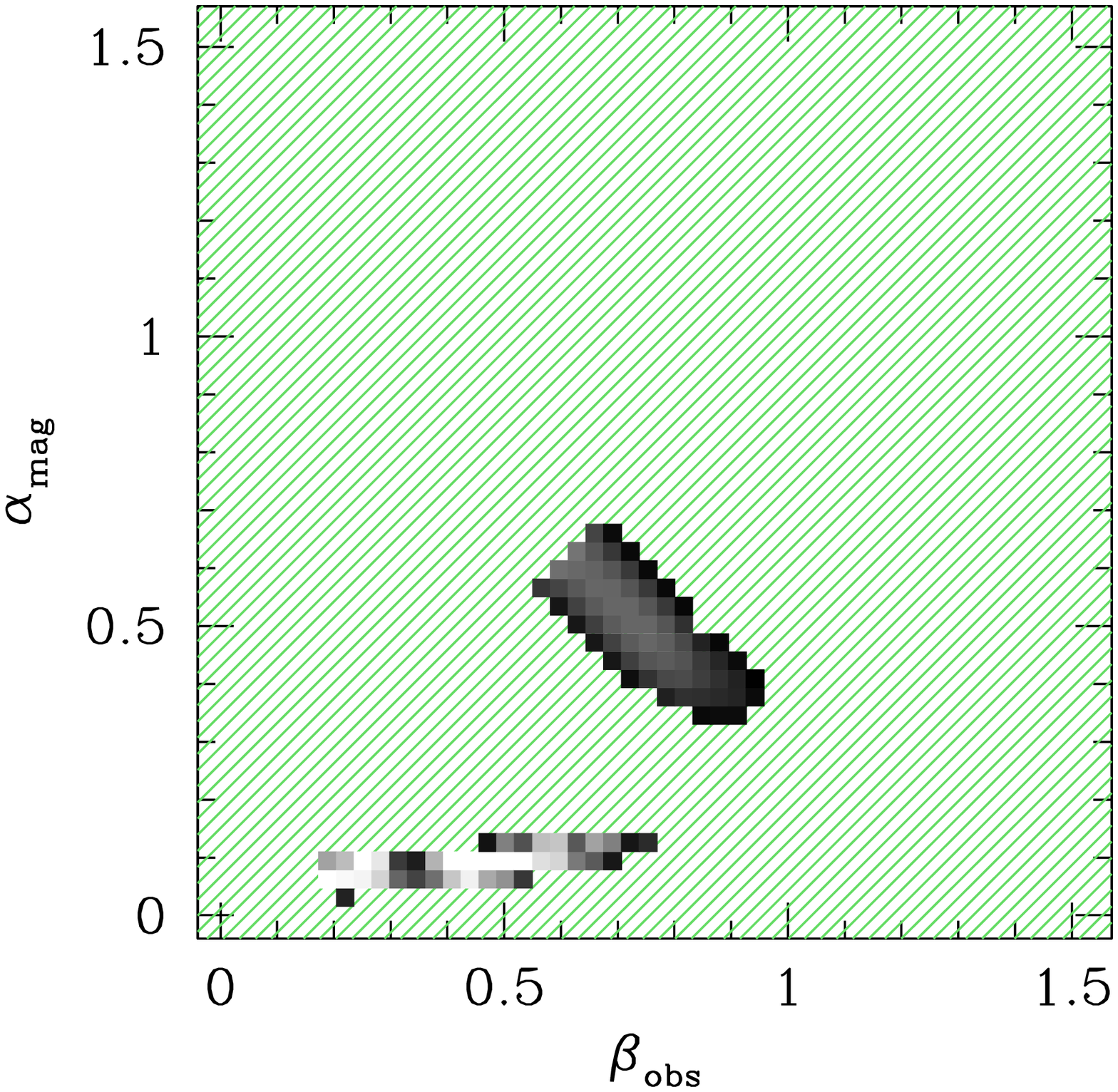} &
\includegraphics[width=0.3\textwidth]{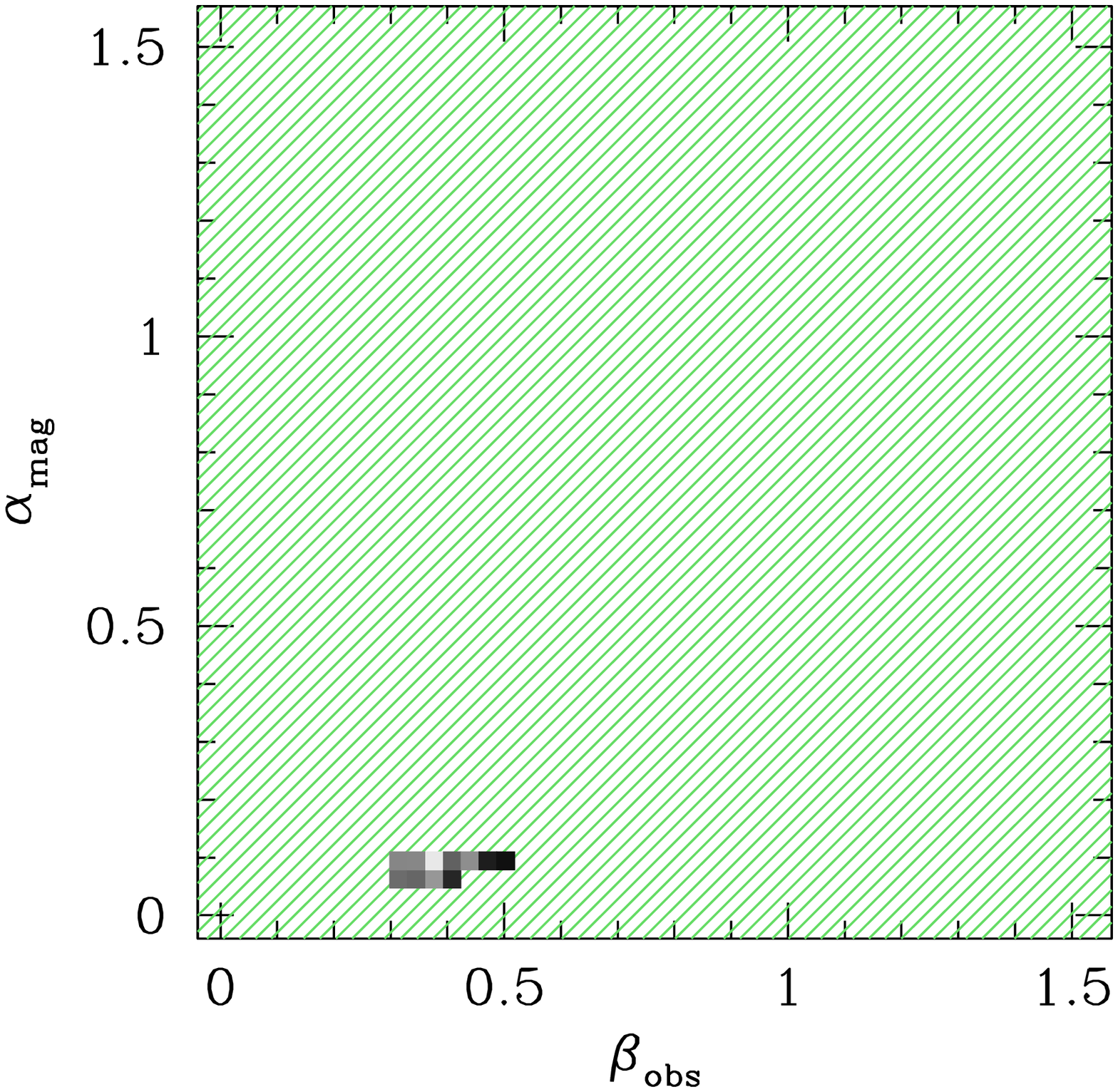}
\end{tabular}
\end{center}
\vspace{-0.8cm}
\caption{
Map of $p$-value for the fit by the coronal outflow model 
for \onee\ (left), \ \fouru \ (middle) and \onerxs  \ (right).
The hatched green region has $p$-values smaller than $0.001$.
}
\label{fig_chi2}
\end{figure}

\fig{fig_chi2} shows the $p$-value maps on the plane of $\alphamag, \betaobs$
after maximizing the $p$-value (minimizing $\chi^2$) over the parameters $\theta_j$ 
and $\lj$ for each $\alphamag, \betaobs$. One can see that the model completely 
fails ($p<0.001$) almost everywhere; this reflects the fact the model is quite ``rigid,'' 
lacking flexibility in adjusting $\chi^2$ to a desired value. Nevertheless, a good 
fit with $p>0.05$  is obtained in a small region of the parameter space.
We conclude that the model (a) successfully fits the data and (b) the fit imposes
strong constraints on the parameters. 

We note, however, that there is one significant source of degeneracy in the 
axisymmetric model. One can show that interchanging the values of $\alphamag$ and 
$\betaobs$ does not change the predicted spectrum. Thus, Figure~2 must be 
complemented with a similar map obtained by mirror reflection about the line of 
$\betaobs=\alphamag$.
The fact that one of the two angles is found to be small for \fouru \ and \onerxs \ may 
help to break this degeneracy.
Since the line of sight is random, there is no reason for $\betaobs$ to be small.
It is more likely that it is $\alphamag$ that is small, as this may be a result of 
the magnetic field evolution in the neutron star at its birth or later times.

\newcommand{\markaa}{\tablenotemark{a}}
\newcommand{\markab}{\tablenotemark{b}}
\newcommand{\markac}{\tablenotemark{c}}
\begin{table}[t]
\begin{center}
\caption{Parameters of the best-fit model for the hard X-ray component
\label{tab_hard_comp}}
\vspace{1.0mm}
\scriptsize{
\begin{tabular}{ccccccc} \hline\hline
           & $\alphamag$ & $\betaobs$ & $\theta_{j}$ & $\lj$\markaa$ @ D_0$\markab & $\mu_{\rm m}$\markac & $\chi^2$/dof \\ 
	 &    (rad)           & (rad)		& (rad)		& 		&   $(10^{32} \ \mathrm{G\ cm^3})$    &		 \\ \hline
\fouru 	 & $< 0.15$ & 0.5(3)	   	  & $<0.23$	 & 3.9(24) @ 3.6   &   1.3   & 46/41  \\	 
\onerxs 	 & $< 0.15$ & 0.4(2)		  & $<0.15$	 & 3.5(18) @ 3.8 	&   4.6   & 49/37  \\	 
\onee \ -- sol. 1	 & 0.3(2) & 0.9(2) & $<0.4$	 & 48(16) @ 8.5 	&  6.9   & 302/267  \\
\onee \ -- sol. 2	 & 0.7(2)	      & 1.4(1)		& $<0.4$	&  48(16) @ 8.5  &    6.9     &326/267  \\ \hline
\end{tabular}}
\end{center}
\footnotesize{{\bf Note.} 
Uncertainties are given at the 3$\sigma$ level.
There are two solutions within 3$\sigma$ for \onee .\\}
$^{\rm a}${In units of $10^{35}(D/D_0)^2\ \rm erg\ s^{-1}$}.\\
$^{\rm b}${Distance to the object (in kpc) estimated from reddening and X-ray extinction 
along the line of sight 
(for \fouru\ and \onerxs , \citealt{durant_2006}) or from the Galactic rotation of the associated supernova remnant (for \onee, \citealt{tian_2008})}.\\
$^{\rm c}${Magnetic dipole moment of the neutron star inferred from its spindown rate (see \citealt{dib_2007} for \fouru , \citealt{dib_2008} for \onerxs \ and \onee ).} \\
\end{table}

We also explored the model with unfrozen parameter $\mu_{\rm m}$ and investigated
if fitting by this more flexible model allows one to obtain constraints on $\mu_{\rm m}$, 
which would be independent from the spindown measurements. We found that such constraints 
are weak ---
only a lower limit on the magnetic moment can be derived this way, which is about one 
order of magnitude below the value inferred from the spindown rate.

\subsection{Soft component}
\label{fit_soft_component}

Once the hard X-ray component is fitted by the $e^{\pm}$ outflow model, 
we turn to the remaining soft component below 10~keV. Note that the outflow
makes a non-negligible contribution below 10~keV (the two components overlap in 
this region). Thus, understanding the origin of the hard X-ray component and its 
low-energy extension is important for the correct interpretation 
of soft X-ray emission. 
Previous models for the 1-10~keV spectrum involved radiative transfer in the 
the neutron star atmosphere (e.g. \citealt{ho_2001, ozel_2001}) or resonant scattering by 
mildly relativistic electrons flowing in the magnetosphere \citep{thompson_2002}.
Alternatively, the soft X-ray tail was proposed to be the signature of a hot spot on the star 
(e.g. \citealt{gotthelf_2007}).

Motivated by the j-bundle picture described in Section~2, we investigated models 
that include the possible presence of a hot spot.
Two-temperature 
blackbody (2BB) is a simplest model of this type. The lower temperature is typically 
associated with a large area (comparable to the area of the neutron star surface) and 
the higher temperature is associated with the footprint of the j-bundle. This simple model 
was sufficient to provide a good fit for the spectrum of \onee\ \citep{an_2013}.
We found that the 2BB model is insufficient for the soft component of \fouru\ and \onerxs. 
This may be expected, as the soft X-ray spectrum should be modified by resonant 
scattering in the magnetosphere (e.g. \citealt{thompson_2002, rea_2008}), 
although a consistent model for 
this effect is yet to be developed (see \citealt{beloborodov_2013} for discussion). 
We chose a phenomenological modification of the 2BB model: we replaced the Wien 
tail of the hotter blackbody by a power law that is smoothly connected to the 
Planck spectrum at some energy $E_{\rm tail}$. Smooth connection here means the 
continuity of the photon spectrum and its first derivative; then $E_{\rm tail}$ is the 
only required additional parameter. The temperatures $T_1$ and $T_2$ of the 
cold and hot blackbodies, their luminosities $L_1$ and $L_2$, and $E_{\rm tail}$ uniquely
define our modified 2BB model.

\begin{figure}[h]
\begin{center}
\hspace*{-1.cm}
\begin{tabular}{ccc}
\onee & \fouru & \onerxs \\
\includegraphics[width=0.34\textwidth]{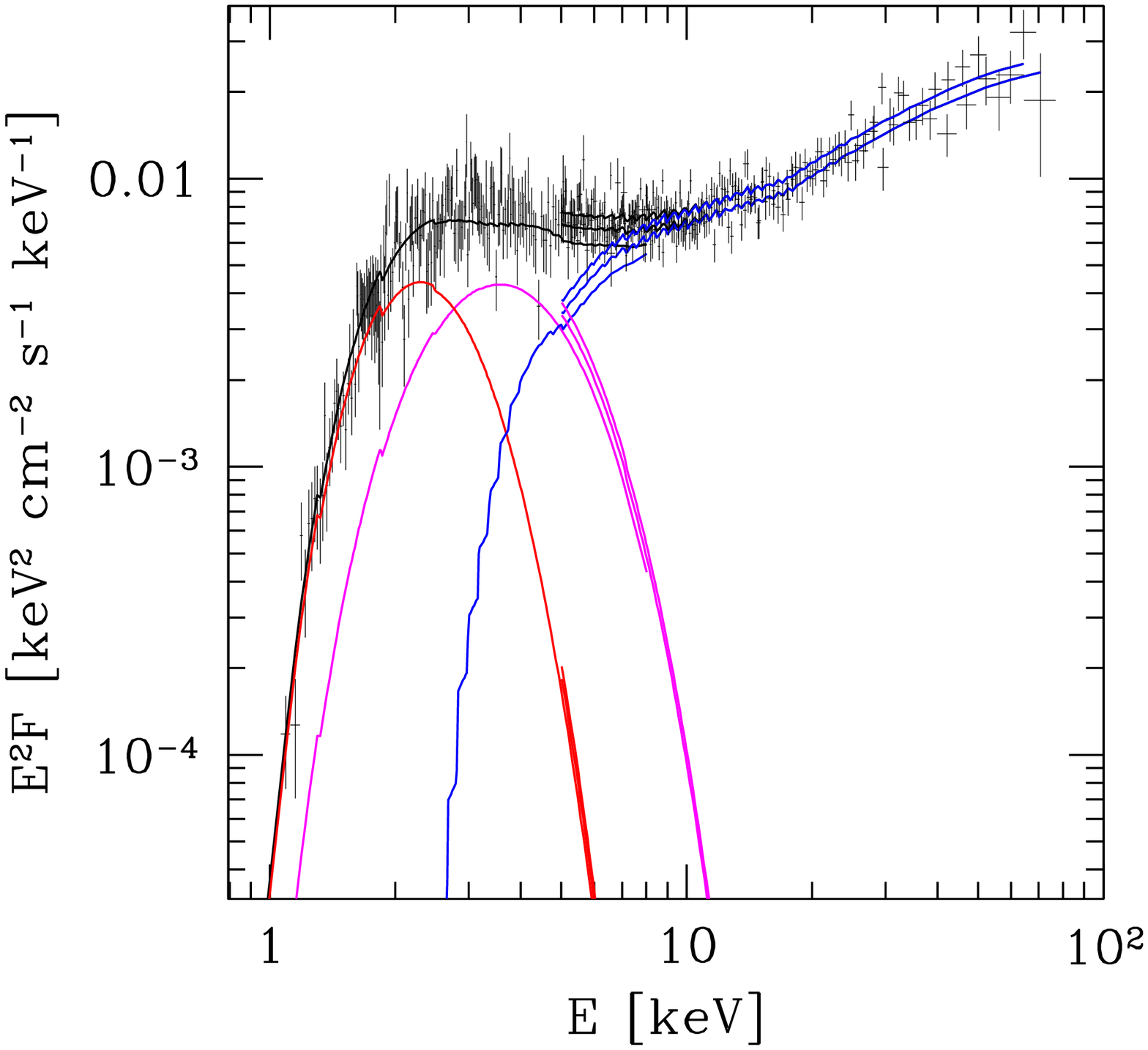} & 
\includegraphics[width=0.34\textwidth]{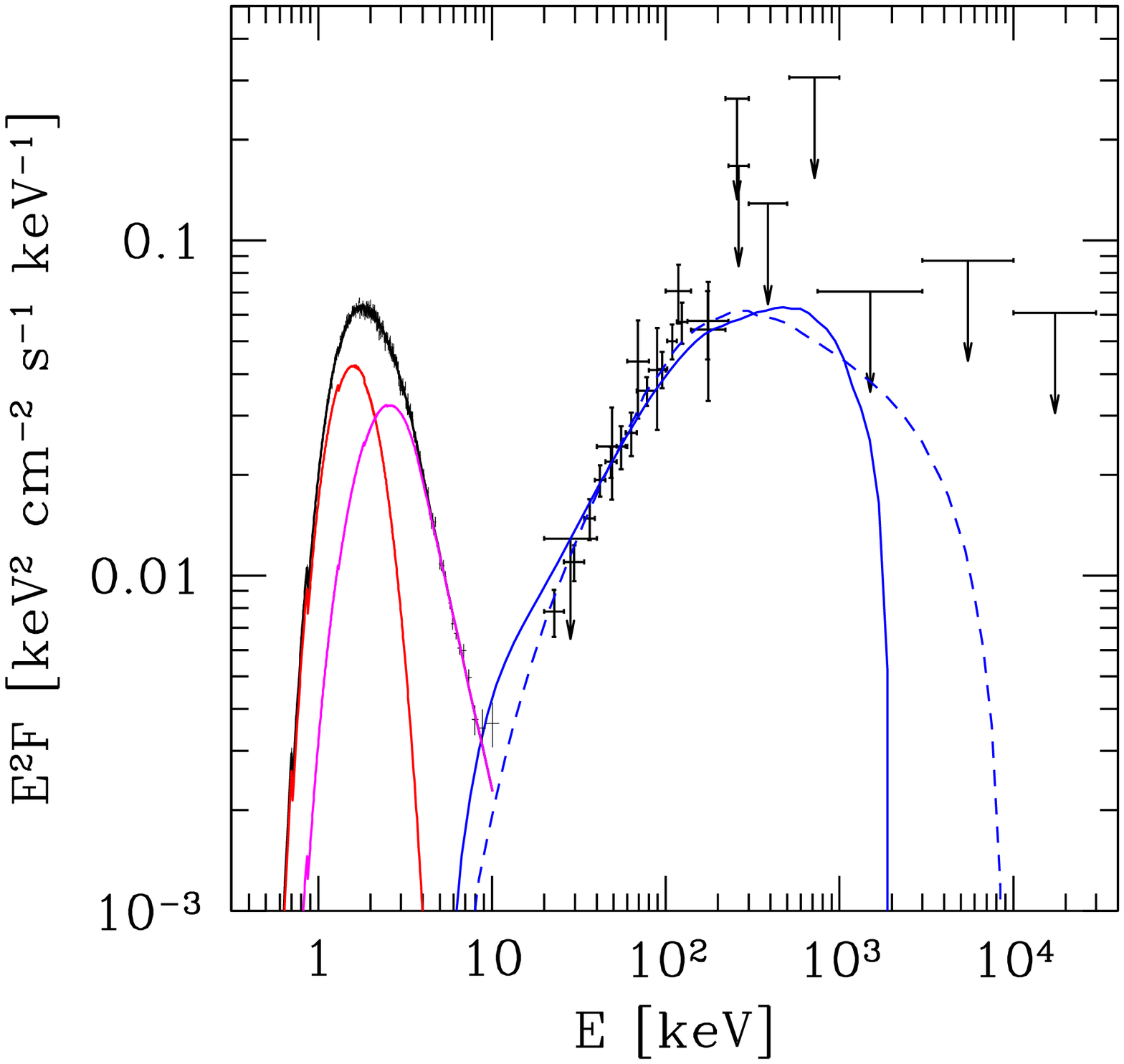} &
\includegraphics[width=0.34\textwidth]{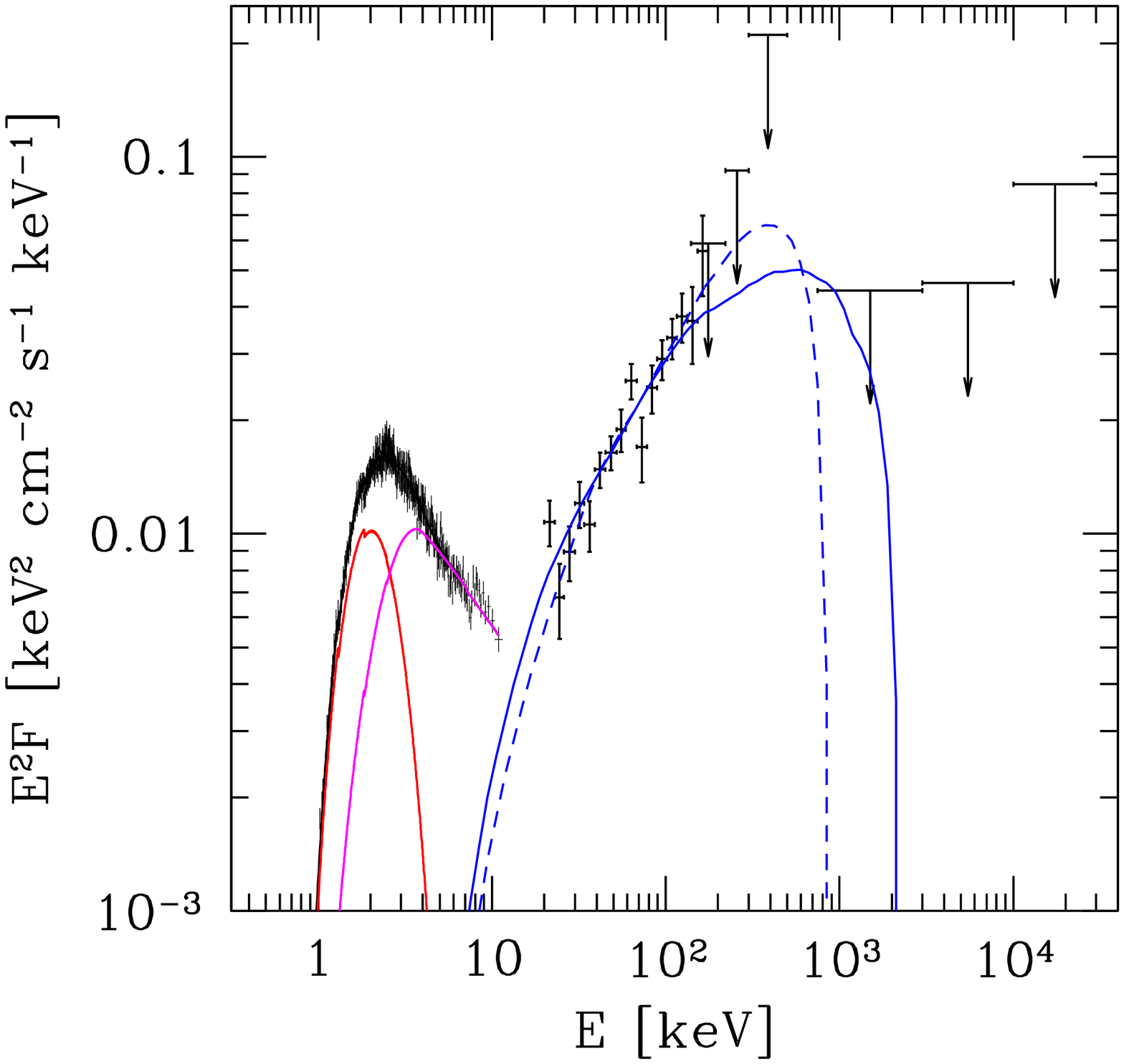}
\end{tabular}
\end{center}
\vspace*{-0.8cm}
\caption{
Best fits for the X-ray spectra of \onee, \fouru, and \onerxs.
Three spectral components are shown in each panel: 
the cold blackbody (red), the hot (modified) blackbody (magenta), and the 
coronal outflow emission (blue).
For \fouru \ and \onerxs, 
the dashed blue curve shows the best fit of the hard component obtained when 
only the phase-averaged spectrum is considered and the phase-resolved data are neglected. 
The data and models in the figure are not corrected for interstellar absorption. 
Only {\it Swift} XRT data are shown for \onee\ 
to avoid confusion in the figure (the actual analysis presented in \citet{an_2013}
includes Chandra and XMM data);
 the model curves split in two because of the offset in the normalization 
between the XRT and the two modules of NuSTAR.
}
\label{fig_spec}
\end{figure}

We found that the spectra of \fouru\ and \onerxs\ are reasonably well fitted by the 
modified 2BB model; the results are summarized in \tab{tab_soft_comp}. 
The formal $\chi^2$ is unsatisfying, because of 
the high spectral resolution of XMM and the inability of simple phenomenological 
models to describe spectral lines.
However the continuum of observed spectra is well reproduced. The residuals are 
largest below 3~keV and suggest the presence of line-like features;
this issue was already pointed by \citet{denhartog_2008}.
The best fits of the soft and hard components are shown together in \fig{fig_spec}.

For comparison, we also tried to fit the soft component with the sum of one (modified) 
blackbody and a power law. This model provided acceptable fits 
for \onerxs \ and \onee , but not for \fouru \ ($\chi^2/$dof$=774/449$).
An additional, well known drawback of this model is 
the large contribution of the power-law below $\sim 1$ keV, which results in an unrealistically large best-fit value of 
the hydrogen column density toward the object.

\newcommand{\markd}{\tablenotemark{a}}
\begin{table*}[t]
\begin{center}
\caption{
Parameters of the best-fit model for the soft X-ray component
\label{tab_soft_comp}}
\vspace{1.0mm}
\scriptsize{
\begin{tabular}{cccccccc} \hline\hline
                  & $N_{\rm H}$      	          & $kT_{1}$ & $kT_{2}$	  &  $E_{\rm tail}$ & $L_1$\markd         &  $L_2$\markd     & $\chi^2$/dof\\ 
                         & ($10^{22}\ \rm cm^{-2}$) 	&  (keV)	   &  (keV)	           &	(keV)         &                                    &	                           &  \\ \hline
 \fouru           &  0.577(5)  & 0.307(4) & 0.602(9) & 3.82(6)  & 2.63(3) & 1.36(9) & 559/448 \\              
 \onerxs        &  0.99(1)    & 0.39(1) & 0.87(5) & 4.0(2) & 0.69(3) & 0.56(12) & 297/228 \\              
 \onee                  &  2.03(4) & 0.45(1) & 0.90(4) & $\cdots$  & 2.15(7) & 0.65(9) & 2298/2272 \\              
\end{tabular}}
\end{center}
\footnotesize{{\bf Note.} }
$^{\rm a}${ In units of $10^{35}(D/D_0)^2\ \rm erg\ s^{-1}$. See \tab{tab_hard_comp} for the values of $D_0$.}\\
\end{table*}

\section{Discussion}
\label{discussion}

Remarkably, the coronal outflow model fits the data in a small region of the 
parameter space, which allowed us to derive significant constraints on the active
bundle of magnetic field lines (j-bundle).
In particular, for \fouru \ and \onerxs\ 
the opening angle of the j-bundle $\theta_j\sim 0.1-0.2$, and its total luminosity 
(most of which is emitted in the MeV band, outside the observed spectral range) is 
$\sim 4\times 10^{36}$~erg~s$^{-1}$. Both parameters are in line with theoretical 
expectations \citep{beloborodov_2009}.
Using Equation~(3) 
one can estimate the product of the twist angle $\psi$ and the discharge voltage 
$\voltage$ in the j-bundle. We found $\psi\,\voltage\simgt (2-3)\times 10^{10}$~V,
consistent with $\voltage\simgt 10^{10}$~V.\footnote{
   The twist angle $\psi\simgt 1$ is expected for an active magnetar. The growth of $\psi$ 
   beyond a critical value $\sim 3$ is prohibited by a global instability --- 
   the over-twisted magnetopshere inflates, forms an unstable current 
   sheet, and ejects magnetic plasmoids, reducing the twist energy \citep{parfrey_2013}.}  
This voltage is higher
(at least by a factor of a few) compared with theoretical estimates of \citet{beloborodov_2007}
or the voltage inferred from the outburst decay in XTE~J1810$-$197 
\citep{beloborodov_2009}.

The j-bundle parameters obtained for \fouru \ and \onerxs \ are comparable, 
which may be expected, as their spectra are similar. The spectrum of \onee \ is 
significantly different: the spectral index of its hard X-ray component is larger and 
the dip between the soft and hard components is smaller. 
We find that this difference is mainly 
explained by different angles $\alphamag$ and $\betaobs$; the parameters 
of the j-bundle itself ($\lj$ and $\theta_j$) are comparable for all three objects.
The fact that two out of three magnetars appear to have a small angle 
between the magnetic and rotation axes $\alphamag\sim 0.1$ suggests that rotation 
plays a role in shaping the magnetic moment of the neutron star.
Future measurements of phase-resolved spectra for other magnetars 
will allow one to better study the statistics of $\alphamag$.
Independent constraints on $\alphamag$ and $\betaobs$ can be provided by 
measurements of X-ray polarization by future missions 
(such as Astro-H, \citealt{takahashi_2012}).

The analysis of the hard X-ray component 
gives the magnetic flux in the j-bundle from which one can estimate its footprint 
area ${\cal A}_j$. The exact area and location of the footprint depend on 
details of the magnetic field near the star; a rough estimate for ${\cal A}_j$ may 
be obtained assuming a dipole field. A hot spot with area $\sim {\cal A}_j$ may be expected on the star.
The obtained fits of the soft X-ray component are indeed consistent 
with the presence of a hot spot and give an independent estimate for its area
${\cal A}_2$. In \onerxs\ and \onee, ${\cal A}_2$ is consistent with ${\cal A}_j$,
providing further support to our model.
For \fouru\ we found ${\cal A}_2\approx 6{\cal A}_j$, suggesting a hot spot extending around the 
j-bundle footprint or indicating that the centered dipole configuration 
is a poor approximation to the magnetosphere near the star.

Instead of studying the phase-resolved spectra in three phase bins, one could use 
coarse spectral binning and analyze more detailed pulse profiles.
We did not attempt to fit the detailed pulse profiles, because they are significantly 
affected by moderate deviations from axisymmetry.\footnote{
The j-bundle emission is relativistically beamed along the magnetic field lines,
especially in the hard X-ray band. Breaking the axial symmetry 
(concentration of electric currents in a range of magnetic longitude) 
affects the observed pulse profiles.} 
Relaxing the axisymmetric assumption would make the model more flexible and  
difficult to describe using a small number of parameters. This would require a 
different, more complicated fitting procedure.
Note also that the pulsed fractions of the three magnetars are moderate even at high 
energies, in the range of 30-40\% at $\sim 100$ keV (W. Hermsen, private communication).
This is consistent with approximate axisymmetry of the j-bundle, 
explaining why our simple model gives reasonable fits to the phase-resolved spectra
in wide phase bins.

The coronal outflow model predicts that most of the luminosity is emitted 
in the MeV band. Future detectors with sensitivity better than that of Comptel
would help to better constrain the location of the j-bundle and the angle 
between the magnetic and rotation axes.

\acknowledgements
This work was supported by NASA ATP grant  NNX 13AI34G.

\bibliographystyle{apj}


\end{document}